\documentclass[aps,prd,superscriptaddress,twocolumn,nofootinbib,10pt]{revtex4-2}
\usepackage{amsmath}
\usepackage{amssymb}
\usepackage{amsthm}
\usepackage{natbib,aasdefs,url,bm}
\usepackage{array}
\usepackage{float}
\usepackage{graphicx}
\usepackage{subfigure}
\usepackage{color}
\usepackage{booktabs}
\usepackage{ctable}
\usepackage{threeparttable} 
\usepackage{CJK}

\usepackage{tablefootnote}

\newcommand{\btheta}{{\boldsymbol{\theta}}}
\long\def\exclude#1{}

\newcommand{\Phinorm}{{(E_\nu^2 \Phi_{\nu+\overline{\nu}}^{\rm per \,flavor})_{20\,\rm TeV}}}

\newcommand{\equ}[1]{Eq.~(\ref{equ:#1})}
\newcommand{\figu}[1]{Fig.~\ref{fig:#1}}

\newcommand{\uheevent}{KM3-230213A}

\def\hlinewd#1{
\noalign{\ifnum0=`}\fi\hrule \@height #1 
\futurelet\reserved@a\@xhline}

\begin{document}

\title{Single-source-class interpretation of the diffuse astrophysical neutrino flux}

\author
{Walter Winter}
\affiliation{Deutsches Elektronen-Synchrotron DESY, 
Platanenallee 6, 15738 Zeuthen, Germany}

\author
{Damiano F.\ G.\ Fiorillo}\affiliation{Istituto Nazionale di Fisica Nucleare (INFN), Sezione di Napoli, Complesso Universitario di Monte Sant'Angelo, Via Cintia, 80126 Napoli, Italy}

\author
{Sara Buson}\affiliation{Deutsches Elektronen-Synchrotron DESY, 
Platanenallee 6, 15738 Zeuthen, Germany}
\affiliation{Julius-Maximilians-Universit{\"a}t W{\"u}rzburg, Fakult{\"a}t f{\"u}r Physik und Astronomie, Institut f{\"u}r Theoretische Physik und Astrophysik, Lehrstuhl f{\"u}r Astronomie, Emil-Fischer-Stra{\ss}e 31, 97074 W{\"u}rzburg, Germany}

\begin{abstract}
We explore the interpretation that the diffuse astrophysical neutrino flux is dominated by a single standard candle-like source class. Since recent observations favor a broken power law with a spectral break around 30~TeV, we postulate that the $p\gamma$ channel is the dominant neutrino production process creating a peak at these energies. We use a SOPHIA-based photo-pion interaction model with a thermal target including high-energy processes, such as multi-pion production, which turns out to be relevant for the interpretation. We demonstrate that target photon temperatures of about 80 to 300 eV are preferred in a multi-parameter fit, whereas the maximal neutrino energies can be limited by A) soft injection spectra, B) a maximal proton energy in the PeV range, or C) magnetic field effects on the secondary muons, pions, and kaons with $B$ in the few 10 kG range.  We predict that future measurements, such as of the neutrino flavor composition or neutrino-antineutrino ratio (Glashow resonance), can discriminate scenarios. We also point out that the parameters obtained in our generic approach, such as in the strong magnetic field values, might be indicative for an AGN core origin as a driver of the diffuse flux.
\end{abstract}

\maketitle

\section{\label{sec:intro}Introduction}

Ever since a diffuse TeV-PeV flux of astrophysical neutrinos has been discovered \citep{IceCube:2013low}, the analyses of different samples from the IceCube collaboration have routinely followed the power-law template for the diffuse neutrino spectrum: examples are recent muon track \citep{Abbasi:2021qfz,IceCube:2024fxo} and cascade \citep{IceCube:2020acn} analyses.
Even though alternative hypotheses, such as log parabola or broken power law, have been tested (see e.g. the cited muon track and cascade analyses), only very recently the evidence for a spectral feature has been emerging, which is a spectral break at about 30~TeV~\citep{IceCube:2025tgp}. Alternative empirical models, such as a log-parabola or a muon-damped descriptions, seem to describe data similarly well \citep{IceCube:2025ewu}.
The fitted spectral models are, however, not self-consistent spectra obtained from neutrino production calculations, but rather empirical assumptions for the neutrino spectral shape and flavor composition. 

The recent observation of a neutrino with a likely neutrino energy around 220~PeV by the KM3NeT collaboration~\citep[\uheevent,][]{KM3NeT:2025npi} further complicates this picture, being in slight tension with the  upper limits from IceCube and Auger at high energies~\citep{KM3NeT:2025ccp}. 
Extrapolating the diffuse spectrum at the highest energies with a power-law suggests that \uheevent\  is consistent with an upward fluctuation, whereas a spectral hardening around PeV energies --- as expected from an additional population of sources --- could not be statistically established~\citep{KM3NeT:2025ccp}.
As a consequence, overall neutrino data point towards a spectral break at low TeV energies, followed by a power law at higher energies, which could in principle be generated from a single (dominant) population of sources. Note, however, that there are arguments in terms of specific source classes which point towards multiple competing populations of sources to the diffuse neutrino flux dominating in different energy ranges, see e.g. \citet{Palladino:2018evm}. 
Moreover, evidence for several sub-leading contributions to the diffuse neutrino flux has been reported, such as from the Galactic plane~\citep{IceCube:2023ame}, the active galaxy NGC 1068~\citep{IceCube:2022der}, and the blazar TXS 0506+056~\citep{IceCube:2018cha}. None of these individual contributions reaches the level of the total diffuse flux in any energy range, unless extrapolated to an entire source population. It therefore remains unclear whether any of these populations powers the bulk of the diffuse neutrino emission, or whether different populations dominate in different energy bands.
We focus on the hypothesis that a single population dominates across the entire neutrino spectrum, which is the simplest scenario still compatible with these observations.  

Self-consistently produced neutrino fluxes and flavor compositions have been studied to describe data earlier e. g. for $p\gamma$ (proton-photon) \citep{Winter:2013cla} and $pp$/$Ap$ (proton-proton/nucleus-proton) \citep{Winter:2014pya} interactions.
These models typically rely on few assumptions, such as in a model for $p\gamma$ interactions in \citet{Hummer:2010ai}:
a proton population injected with a power-law spectrum, target photons as synchrotron radiation from a population of co-accelerated electrons, certain magnetic fields in the source, and a size of production region. The maximal proton energy, which the maximal neutrino energy typically follows, can be self-consistently determined under certain assumptions from the acceleration and dominant cooling rates. In light of the moderate model-dependencies, a major simplification comes from \citet{Fiorillo:2021hty}, where it was demonstrated  that the neutrino spectrum, even from non-thermal photon spectra, can be well approximated by assuming a thermal target photon spectrum with a suitably chosen effective temperature. Note that this can be a black body or grey body spectrum as we will not specify its normalization, see below. 
Owing to the scenario of a single-population dominated diffuse flux, we assume that neutrinos are produced by $p\gamma$ interactions in this work. As we will show, this production channel can naturally produce a smooth bump in the TeV range, providing a natural explanation for the observed spectral break. In contrast, $pp$ interactions would require a tailored-shaped low-energy cutoff of the non-thermal proton population to mimic this effect, which might be less appealing.

In order to describe the neutrino spectral shape, flavor and neutrino-antineutrino composition accurately, several ingredients are needed: first, neutrino production channels beyond the $\Delta$-resonance have to be considered, such as in the SOPHIA software~\citep{Mucke:1999yb}. For example, the multi-pion channels will lead to power-law neutrino spectra even for thermal target photons if the proton energies are high enough \citep{Fiorillo:2021hty}; this effect is crucial in order for thermal target photons to mimic the effect of a non-thermal target. Second, secondary muons, pions and kaons may cool in strong magnetic fields, leading to spectral and flavor distortions at high energies~\citep{Kashti:2005qa,Lipari:2007su}. This effect is especially prominent in sources such as gamma-ray bursts  \cite[e.g.][]{Baerwald:2010fk}, microquasars~\citep{Reynoso:2008gs}, as well as in the coronae of active galactic nuclei (AGN) if they are powered by magnetic reconnection~\citep{Karavola:2024uui,Karavola:2026rpg}. Note that the kaon production channel may then dominate at the highest energies~\citep{Kachelriess:2007tr}. 

Our approach produces realistic neutrino spectra and flavor compositions from $p\gamma$ interactions, but does not address the normalization -- which depends on additional parameters, such as proton luminosity, optical thickness to $p\gamma$ interactions and source density distribution. In this sense, we are taking an agnostic viewpoint with respect to any astrophysics-motivated source class, focusing rather on the general question: can the neutrino spectral break be described by a single source class and, if so, what source physical properties could produce it? 
We will also discuss how predictions for future instruments may discriminate different degenerate solutions. Our approach fills the gap between empirical fit models used by the experimental collaborations and more sophisticated theoretical astrophysical source models, which require more ingredients.

We structure the paper as follows: in Sec.~\ref{sec:methods}, we present the main properties of the thermal model adopted, discuss the neutrino production, and describe the statistical analysis we perform. In Sec.~\ref{sec:fitresults}, we collect the results of this analysis, discussing the favored explanations for a spectral break, and in Sec.~\ref{sec:predictions} we consider possible future observables which might help discriminate between these different explanations for the break. Finally, in Sec.~\ref{sec:discussion} we discuss our results in the wider context and the limitations of our approach, and we summarize our conclusions in Sec.~\ref{sec:summary}. Throughout this work, we use Gaussian units as conventional in astrophysics, where $c$ is the speed of light, but the Boltzmann constant is assumed to be $k_B=1$ (temperatures in $eV$).

\section{Methods and data used}
\label{sec:methods}

We use the NeuCosmA software to simulate realistic neutrino spectra \citep{Hummer:2010vx,Hummer:2011ms,Biehl:2017zlw}, which uses a photohadronic production method based on SOPHIA~\citep{Mucke:1999yb}. The $p\gamma$ interactions are based on \citet{Biehl:2017zlw}, using a slightly updated and more accurate method than \citet{Hummer:2010vx}. We however checked that, for the model used in this work, the differences are small (see also \citet{Cerruti:2024lmj} for a comparison of the outcome of different neutrino production codes).

For our generic model, we assume a population of standard candle sources at redshift $z \simeq 1$ with an isotropically emitting production region, which moves with a negligible Doppler factor towards the observer. A potential Doppler boost would, in principle, affect the normalization and the energy of the produced neutrinos, an effect that is degenerate with the other parameters of the model.  
The chosen redshift comes from the observation that the main contribution to the diffuse neutrino flux from any source population following the star formation rate will predominantly come from $z \simeq 1$. This does not limit the applicability of our model, as we checked that the impact of using such a redshift distribution is small.\footnote{The dominance of $z \simeq1$ can be seen by convolving typical star formation rates $F(z)$, see e.g. \citet{Hopkins:2006bw}, with volume correction $dV/dz$ and $1/(4 \pi d_L^2)$ weight factors. We checked that in this case the effect on our neutrino spectra is small, even if additional powers of $(1+z)$ are included, such as for a transient population. Slightly stronger deviations may be expected from sources distributed very differently, such as with negative source evolution or high redshifts only.
}

We assume that protons are injected from an acceleration zone with a differential (in energy) proton spectrum
\begin{equation}
    N_p(E_p) \propto E_p^{-\alpha_p} \, \exp \left( - \frac{E_p}{E_{p,\mathrm{max}}} \right) \, .
    \label{equ:pinj}
\end{equation}
These protons will interact with a thermal target spectrum with temperature $T$ to produce secondary $\pi^+$, $\pi^-$, $K^+$ (leading kaon channel taken into account from \cite{Hummer:2010vx}; see also \cite{Lipari:2007su}, where different kaon production channels are quantitatively compared), and neutrons in the optically thin (to $p\gamma$ interactions) regime. This means that protons interact at most once, and secondary neutrons, which are electrically neutral, can escape the source and will decay into neutrinos via the channel $n \rightarrow p + e^- + \bar \nu_e$ on their path towards Earth. Note that in optically thick sources, the $\pi^-$ to $\pi^+$ ratio can be strongly enhanced, which can affect observable signatures such as the Glashow resonance~\citep{Biehl:2016psj}. 

In practice, the assumed in-source spectrum in \equ{pinj} corresponds to the injection spectrum if the dominant radiation processes are energy-independent, such as adiabatic cooling, free-streaming escape or the dynamical timescale. Energy-dependent radiation processes will have an imprint on the proton acceleration spectrum, and, similarly, different  specific acceleration mechanisms, such as magnetized reconnection~\citep{Zhang:2021akj,Zhang:2023lvw,Chernoglazov:2023ksi,Fiorillo:2023dts}. These could offer alternative explanations for the neutrino spectral break~\citep{Karavola:2026rpg}, as we will discuss in Sec.~\ref{sec:discussion} -- but can be effectively described in our model, as we will demonstrate. A spectral break in the proton spectrum can also arise in optically thick (to $p\gamma$ interactions) sources. Since such a break appears at the $p\gamma$ threshold energy, it just changes the interpretation of $\alpha_p$ as in-source (instead of injection) spectral index and does not introduce additional spectral features.

Pions will decay into muons, where we distinguish four species, $\mu^+_L$, $\mu^+_R$, $\mu^-_L$, $\mu^-_R$, to account for helicity-dependent muon decays~\citep{Lipari:2007su}. The muons then decay further into neutrinos, which can escape from the source. We take into account that the (charged) secondary muons, pions, and kaons, beyond a species-dependent critical energy (see below), may lose energy faster than they decay by synchrotron losses in magnetic fields $B$. We therefore produce all secondaries explicitly and use steady-state solvers for the secondary cooling to maintain accurate flavor ratios through the neutrino production chain, assuming that the cooling and decay is faster than the dynamical timescale of the system or adiabatic cooling, if applicable.\footnote{NeuCosmA can solve the coupled differential equation system for all species both with a time-dependent and steady-state method. For observables on linear scales, such as flavor ratios, the time-dependent implicit solvers produce less accurate results. We therefore use the steady-state method here.} Overall, neutrinos are produced from the secondary pions, muons, kaons or neutrons, determining their spectral shape to leading order, see e. g. App.~A in \citet{Baerwald:2011ee} for a detailed example for gamma-ray bursts.

\textbf{We consequently use $\boldsymbol{\alpha_p}$, $\boldsymbol{E_{p,\mathrm{max}}}$, $\boldsymbol{T}$ and $\boldsymbol{B}$ as our main parameters}. Similarly to many AGN blazars, we parameterize the maximal proton energy instead of deriving it self-consistently -- which would require additional assumptions about acceleration mechanism, target photons, and size of the production region (e. g. for the confinement condition). 
For the same reasons, we also do not consider the co-produced gamma-rays explicitly, as in-source effects may lead to an electromagnetic cascade, see e. g. ~\citet{Fiorillo:2025kuh}, which depends on additional parameters such as the optical thickness to $\gamma \gamma$ interactions. The propagation in the extragalactic background light can also affect the spectrum in a redshift-dependent way. Furthermore, the thermal model for the target photon spectrum, which we use here, is only meant to accurately reproduce the neutrino emission~\citep{Fiorillo:2021hty}, but does not describe the electromagnetic spectrum self-consistently or completely.

The properties of the neutrino emission associated with this model have been extensively studied in~\cite{Fiorillo:2021hty}, so we do not entertain a full model characterization, but rather summarize the results of that study here. For a photon temperature $T$, the neutrino spectrum is always rather hard below a typical energy $E_{\nu,\rm thr}\sim 0.25 \, y_\Delta \, \chi_{p \rightarrow \pi} \, m_p c^2/(3T)$, where $y_\Delta\simeq 0.2\,\mathrm{GeV}$, $m_p$ is the proton mass, $\chi_{p \rightarrow \pi}\simeq 0.2$, and we are assuming that most target photons are at a characteristic energy $\simeq 3 \, T$. Hence, the interaction with the target photons naturally imprints a spectral break at the characteristic energy
\begin{equation}
    E_{\nu, \rm thr}\sim 100\,\mathrm{TeV}\, \frac{100 \, \mathrm{eV}}{3 \,T} .
    \label{equ:ET}
\end{equation} 
At higher neutrino energies,  thanks to the multi-pion contribution to $p\gamma$ interactions, neutrinos roughly follow the proton spectrum, so they continue as a power law with a spectral index $\simeq \alpha_p$ as the protons up to a maximal proton energy $E_{\nu, \rm max}\simeq 0.05 E_{p,\rm max}$. These conclusions can be affected by sufficiently strong magnetic fields, which cool muons and pions by synchrotron emission; their impact (from muon cooling) is felt  above the critical neutrino energies $E_{\nu}^{\mu,\rm damp}\simeq 10^6\,\mathrm{GeV} \, (B/1\,\mathrm{kG})^{-1}$. At an order of magnitude larger energy $E_{\nu}^{\pi,\rm damp} \gtrsim 10^{7}\,\mathrm{GeV} \, (B/1\,\mathrm{kG})^{-1}$, even pions are damped. If the magnetic fields are extremely large ($\gg \mathrm{kG}$), neutrinos from neutron decays dominate because neutrons (as electrically neutral particles) do not lose energy by synchrotron emission; we will discuss this possibility in Secs.~\ref{sec:fitresults} and~\ref{sec:predictions}. 

The neutrino flux obtained through this procedure is compared with the observed diffuse neutrino spectrum, focusing in particular on the recent IceCube results exhibiting a spectral break~\citep{IceCube:2025tgp}. Here the IceCube collaboration reports the inferred neutrino flux from two different data samples: the combined fit (CF) sample based on a combination of track and all-sky contained cascade events, and the medium energy starting events (MESE) sample, which includes only starting events with the neutrino-nucleon interaction vertex within the detector, extending to energies lower than 10~TeV. A direct analysis of these data samples is not feasible without a detailed description of the experiment response function for each data sample. However, we can indirectly compare our model with the observations using the inferred diffuse neutrino flux in different energy bins, using the segmented flux reported in Fig.~3 of~\cite{IceCube:2025tgp}. Thus, for each choice of our model parameters, we extract the integrated energy flux in each of the energy bins of the segmented fit from the IceCube collaboration. The parametric set associated with our model will be denoted by $\btheta=\left\{\alpha_p, E_{p,\rm max},T,B,\Phinorm\right\}$, where the latter parameter denotes the value of the diffuse neutrino flux $E_\nu^2 \Phi_{\nu+\overline{\nu}}^{\rm per\, flavor}$ summed over neutrinos and antineutrinos and divided by three, evaluated at a characteristic energy $E_\nu=20\,\mathrm{TeV}$, close to the inferred spectral break. We will therefore denote the neutrino energy flux by $\Phi_\nu[E_\nu, \btheta]$.

If we call the edges of the $i$-th energy bin, as reported in~\cite{IceCube:2025tgp}, by $E^i_{\nu,\rm min}$ and $E^i_{\nu,\rm max}$ respectively, we can determine the energy flux level (measured in GeV~cm$^{-2}$~s$^{-1}$~sr$^{-1}$) in every energy bin as

\begin{equation}
   \mathcal{F}_i[\btheta]=\frac{\int_{E^i_{\nu,\rm min}}^{E^i_{\nu,\rm max}} E_\nu^2 \Phi_\nu[E_\nu \btheta]d\log E_\nu}{\log(E^i_{\nu,\rm max}/E^i_{\nu, \rm min})}.
\end{equation}
This energy flux is assumed to be per flavor and summed over neutrinos and antineutrinos, and can therefore directly be compared with the segmented energy flux reported in~\cite{IceCube:2025tgp}. In the main text, we focus on the CF data sample, while we discuss in App.~\ref{sec:mese} how the results change with the MESE sample. In both cases, we extract from Fig.~3 of~\cite{IceCube:2025tgp} the measured flux in each bin $\overline{\mathcal{F}}_i$, together with its upper $\sigma_{i,+}$ and lower $\sigma_{i,-}$ uncertainty at $1\sigma$ confidence level. For a few energy bins, only an upper bound on the flux $\mathcal{F}_{i,\rm up}$ is available. To compare these measurements with our model predictions, we construct a log-likelihood (we already multiply it by $-2$, as must be done to obtain a proper chi-squared variable)
\begin{align}
    -2\log\mathcal{L}[\btheta] = & -2\sum_{i,\rm data}\log\mathcal{L}_{\rm data}[\mathcal{F}_{i}[\btheta]; \overline{\mathcal{F}}_i,\sigma_{i,+},\sigma_{i,-}] \nonumber \\ & -2\sum_{i,\rm up}\log \mathcal{L}_{\rm up}[\mathcal{F}_i[\btheta];\mathcal{F}_{i,\rm up}].
\end{align}
For the energy bins with a measured flux, we use a bipartite Gaussian likelihood
\begin{align}
    & -2\log\mathcal{L}_{\rm data}[\mathcal{F}_{i}[\btheta]; \overline{\mathcal{F}}_i,\sigma_{i,+},\sigma_{i,-}]= \nonumber \\ & \qquad \frac{(\mathcal{F}_{i}[\btheta]-\overline{\mathcal{F}}_i)^2}{\sigma_{i,+}^2}\Theta(\mathcal{F}_i[\btheta]-\overline{\mathcal{F}}_i)+ \nonumber \\ & \qquad \frac{(\mathcal{F}_{i}[\btheta]-\overline{\mathcal{F}}_i)^2}{\sigma_{i,-}^2}\Theta(-\mathcal{F}_i[\btheta]+\overline{\mathcal{F}}_i),
\end{align}
where $\Theta$ denotes the Heaviside step function. In the definition of the log-likelihood, we are neglecting any normalization factor which is independent of the model parameters and therefore would disappear after minimizing over them.
For the energy bins with an upper bound on the flux, 
we use a log-likelihood inferred from the Poissonian one\footnote{We begin from the Poissonian likelihood for the observed number of events $\mu$; for zero events, as in the case of an upper bound, we have the log-likelihood $-2\log\mathcal{L}_{\rm Pois}[\mu]=2\mu$, so the $68\%$ exclusion level for $\mu$ is given by the condition $\int_0^{\mu_0} e^{-\mu}d\mu=0.68$, i.e. $\mu_0=1.14$. Therefore, we have $-2\log\mathcal{L}_{\rm Pois}[\mu]=2\mu_0 (\mu/\mu_0)=2.28\mu/\mu_0$. Note that the ratio between expected number of events $\mu/\mu_0$ and integrated flux $\mathcal{F}_{i}[\btheta]/\mathcal{F}_{i,\rm up}$ is the same.}
\begin{equation}
    -2\log\mathcal{L}_{\rm up}[\mathcal{F}_i[\btheta];\mathcal{F}_{i,\rm up}]=\frac{2.28\mathcal{F}_{i}[\btheta]}{\mathcal{F}_{i,\rm up}}.
\end{equation}
By this procedure, we can associate with every choice of parameter $\btheta$ a corresponding log-likelihood $-2\log\mathcal{L}[\btheta]$. 

For each of the models, we explore the parameter space using a test statistic (TS) conventionally defined as
\begin{equation}
    \mathrm{TS}[\btheta_\alpha]=2\max_{\btheta_\beta}\log \mathcal{L}[\btheta_\beta]-2\log\mathcal{L}[\btheta_\alpha],
\end{equation}
which according to Wilks' theorem~\citep{Wilks:1938dza} is distributed as a chi-squared variable with four degrees of freedom. The best-fit model is the one that maximizes the likelihood. 

An overall model comparison in the 5-dimensional parameter space is highly expensive from the computational perspective, and not particularly enlightening, due to the strong parameter degeneracies: as we will see, soft proton spectral indices $\alpha_p>2$, low maximal proton energies $E_{p,\rm max}$, and strong magnetic fields $B$, all lead to analogous suppressions in the neutrino flux at the high energies which are difficult to disentangle. Therefore, we focus on two related sub-models as physically separate explanations for the diffuse neutrino flux: a ``small'' magnetic-field model with $B \ll 10 \, \mathrm{G}$ (simulated by fixed $B=0 \, \mathrm{G}$) and $\btheta_{\rm small \, B}=\left\{\alpha_p,E_{p,\rm max},T,\Phinorm\right\}$, and a ``large'' magnetic field model with $B \ge 10 \, \mathrm{G}$, $E_{p,\rm max} \gg 10^9 \, \mathrm{GeV}$ (simulated by $E_{p,\rm max}=10^{11} \, \mathrm{GeV}$), so large that it does not appear in the neutrino flux, and $\btheta_{\rm large \, B}=\left\{\alpha_p,B,T,\Phinorm\right\}$.\footnote{The listed ranges $B \ll 10 \, \mathrm{G}$ and $E_{p,\rm max} \gg 10^9 \, \mathrm{GeV}$ can be inferred from either estimates of where the maximal or critical neutrino energies cannot interfere with the spectral neutrino shape, or the respective corner plots in \figu{corner_noB} and \figu{corner_B}, we will introduce later, from the parameter ranges where the fit does not change anymore. A specific acceleration mechanism in the same region must roughly respect these constraints.} The two model choices are respectively denoted as $\btheta_\alpha$.

\section{Multi-parameter fit}
\label{sec:fitresults}

\begin{figure*}[t]
    \includegraphics[width=\textwidth]{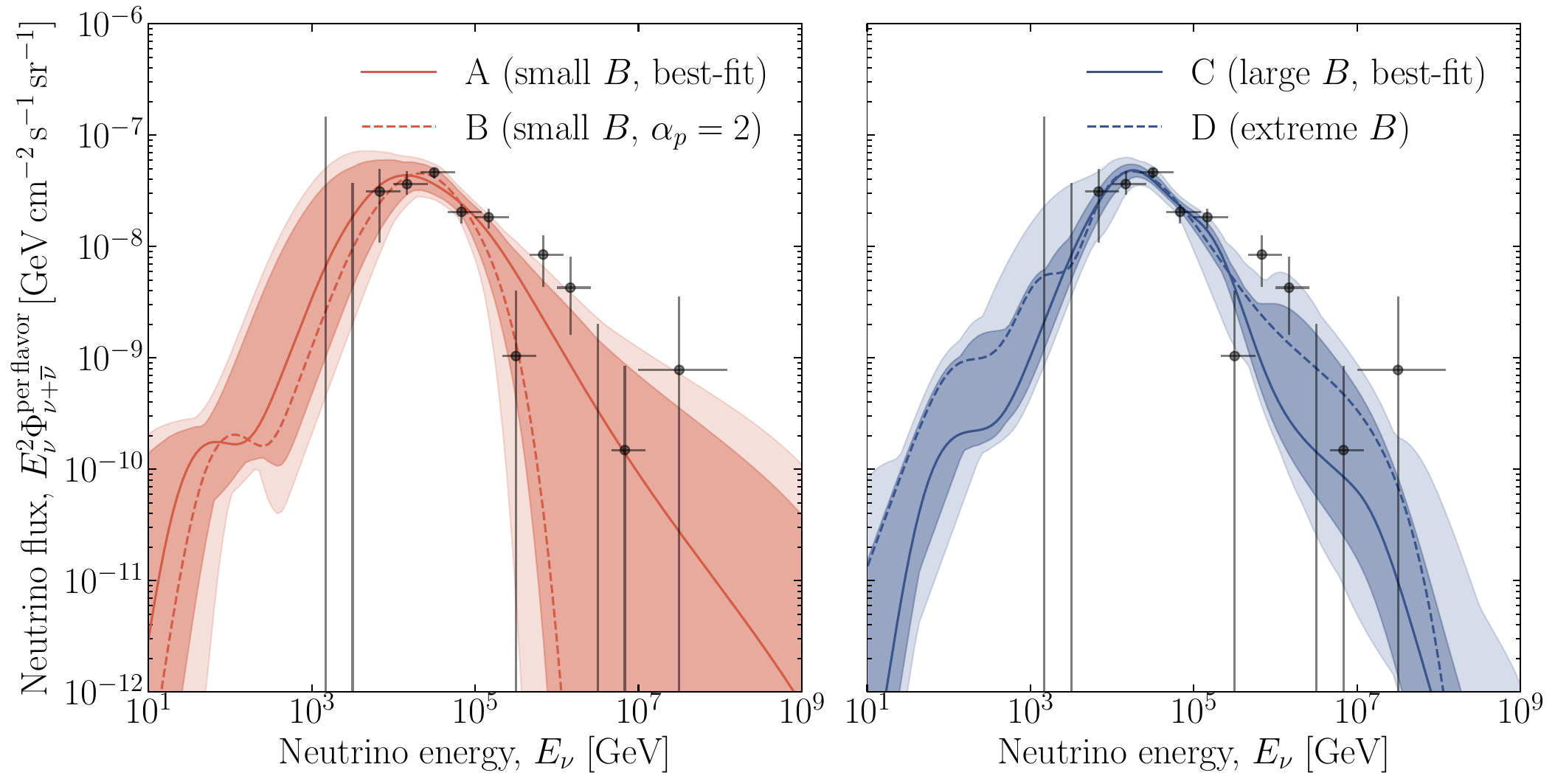}
    \caption{Diffuse neutrino flux spectrum for the models with negligible (left, red) and large (right, blue) magnetic field. Solid curves refer to the overall best-fits for the models (marked as stars in Figs.~\ref{fig:corner_noB} and~\ref{fig:corner_B}). Dashed curves refer to alternative fit models with different physical explanations for the break (denoted as white dots in Figs.~\ref{fig:corner_noB} and~\ref{fig:corner_B}): on the left, we show a model with fixed proton spectral index $\alpha_p=2$ (small $B$ and proton energy cutoff); on the right, we show a model with extreme magnetic field $B=10^7\,\mathrm{G}$. The collected parameters for all four highlighted scenarios are listed in Table~\ref{tab:models}. The colored bands indicate the allowed model fluxes at 68\% and 95\% confidence levels respectively (using 1 d.o.f., corresponding to bin-wise uncertainties). Data points are extracted from Fig.~3 of~\cite{IceCube:2025tgp} (CF).}\label{fig:diffuse}
\end{figure*}

Before explicitly considering the favored regions of parameter space, we directly show the best descriptions for the diffuse neutrino flux in both models (large and small magnetic fields). Fig.~\ref{fig:diffuse} collects the allowed model fluxes. The 68\% and 95\% confidence level bands are obtained by considering the relevant threshold levels for the TS with one degree of freedom. 

\begin{table*}[t]
\centering
\begin{tabular}{lcccc}
\toprule
Scenario & {\bf A} & {\bf B} & {\bf C} & {\bf D} \\
Parameter $\downarrow$ / Description $\rightarrow$ & $~$ Small $B$, best-fit $~$ & $~$ Small $B$, $\alpha_p=2$ $~$ & $~$ Large $B$, best-fit $~$ & $~$ Extreme $B$ $~$ \\
\midrule
$B$ [kG]              & $\ll 0.01$  & $\ll 0.01$  & 56 & $10^4$ \\
$\alpha_p$       & 3 & 2 & 2.6 & 2.4 \\
$E_{p, \rm max}$ [GeV]    & $10^{11}$ & $5.6\times 10^5$ & $\gg 10^9$  & $\gg 10^9$  \\
$T$ [eV]            & 200 & 126 & 100 & 1 \\
$\Phinorm$ [$10^{-8}$ GeV cm$^{-2}$ s$^{-1}$ sr$^{-1}$]         & 4.22 & 4.52 & 4.82 & 4.72 \\
\midrule 
$\chi^2$ & 15.5 & 17.2 & 13.3 & 13.5\\
$\chi^2$/d.o.f. & 1.73 & 1.91 & 1.48 & 1.50 \\
\bottomrule
\end{tabular}
\caption{Collected parameters for the four scenarios used as benchmarks throughout our models. The scanned ranges (priors) for the parameters correspond to the
ranges shown in Figs.~\ref{fig:corner_noB} and~\ref{fig:corner_B}.}
\label{tab:models}
\end{table*}

In both cases, the evidence for a spectral break around 20--30~TeV is clear, as all the allowed models exhibit it precisely around that range. This is driven mostly by the two upper bounds in data below 10~TeV, which require a suppressed flux in that region. We highlight four different explanations for the diffuse neutrino flux, all of which appear as best-fit explanations in selected regions of the parameter space; they exhibit qualitatively different mechanisms for the spectral break, and will therefore be taken as characteristic benchmarks throughout this work. The parameters adopted for each of these scenarios are collected in Table~\ref{tab:models}.

For the scenario with small magnetic field, the best-fit explanation, marked as {\bf scenario A} in Fig.~\ref{fig:diffuse} and in Table~\ref{tab:models}, corresponds to a soft proton flux, with $\alpha_p=3$, and basically no visible cutoff. The spectral break is caused by the target photons with a  favored temperature $T\sim 200\,\mathrm{eV}$ causing the neutrino emission to peak at tens of TeV. The high-energy power-law behavior allows the flux to continue up to tens of PeV energies, providing an explanation even for the highest-energy events, although the data points around 1~PeV are somewhat higher than the model flux prediction.

On the other hand, this high-energy power-law behavior, while favored by the data, is not preferred to a statistically significant level in comparison to an exponentially suppressed flux above the break. To show this explicitly, we have selected a second benchmark, marked as {\bf scenario B}, in which the proton spectral index is fixed to $\alpha_p=2$ (as expected e. g. from diffusive shock acceleration from strong non-relativistic shocks). In this case, the favored photon temperature is still around $T\sim 100\,\mathrm{eV}$, with a neutrino spectral break around 30~TeV; however, the hard neutrino spectrum above the break is cut off by a low maximal proton energy around $E_{p,\rm max}\sim 600\,\mathrm{TeV}$. While the high-energy data points cannot be explained in this scenario, it is still allowed at 68\% confidence level, showing that the limited current statistics do not allow to distinguish yet between high-energy power-law and exponential suppression mechanisms.

\begin{figure*}[tp]
    \includegraphics[width=\textwidth]{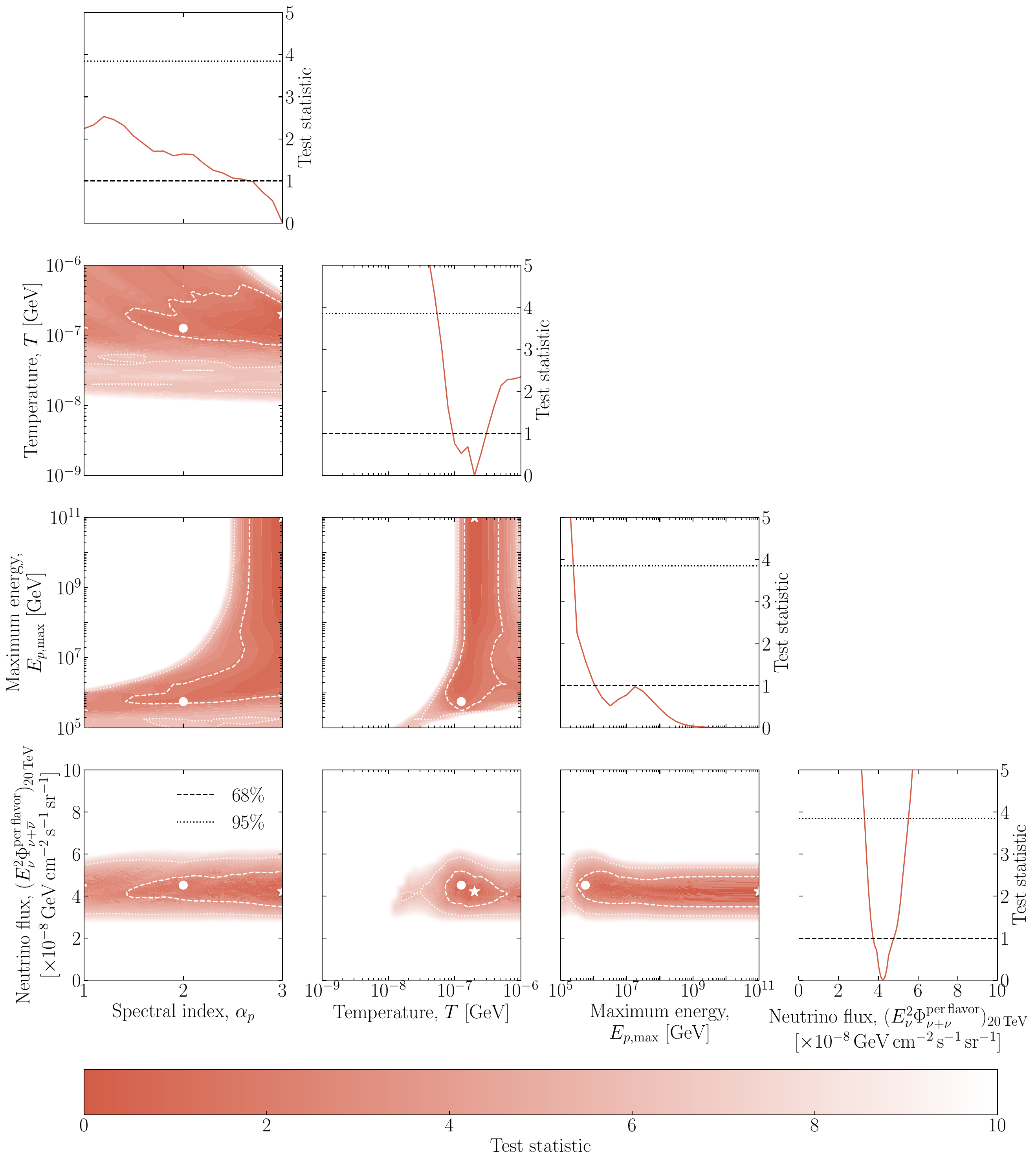}
    \caption{Corner plot for the parameter space of the no-magnetic-field model. The contour plots show the marginalized TS in each pair of parameters, while the corner one-dimensional plots show the marginalized TS for each parameter. The threshold values for TS chosen for two and one degrees of freedom correspondingly. Scenario A (B) is marked by a white star (dot); the best-fit case is scenario A.}\label{fig:corner_noB}
\end{figure*}

\begin{figure*}[tp]
    \includegraphics[width=\textwidth]{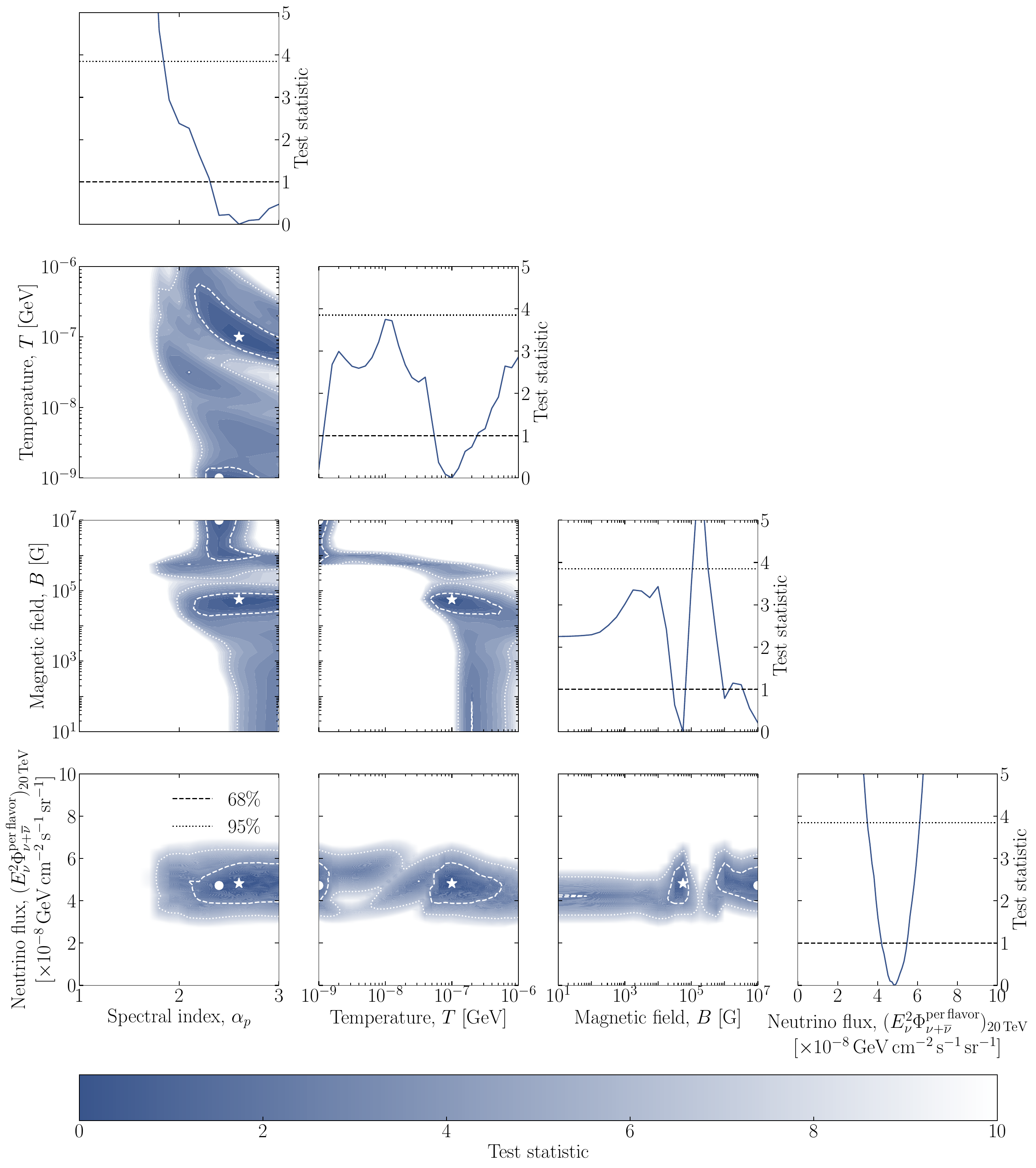}
    \caption{Corner plot for the parameter space of the magnetic-field model. The contour plots show the marginalized TS in each pair of parameters, while the corner one-dimensional plots show the marginalized TS for each parameter. The threshold values for TS chosen for two and one degrees of freedom correspondingly. Scenario C (D) is marked by a white star (dot); the best-fit case is scenario C.}\label{fig:corner_B}
\end{figure*}

An intriguing alternative explanation for the spectral break is instead offered by the scenarios with strong magnetic fields. In this case, the best-fit scenario, marked as {\bf scenario C}, still has a temperature $T\sim 100\,\mathrm{eV}$, but the break is simultaneously accompanied by a muon cooling break due to a magnetic field $B\sim 60\,\mathrm{kG}$. For this reason, the neutrino spectrum has a more irregular structure, coming from the superposition of the neutrino emissions from differently cooled sub-populations of secondaries, namely muons, pions, and kaons, see Sec.~\ref{sec:predictions}.

Finally, in the magnetic field scenario we find a second likelihood maximum {\bf scenario D}, comparable with the best-fit one, corresponding to an extremely strong magnetic field $B=10^4\,\mathrm{kG}$, at the edge of the parameter range we choose. In this case, the photon energy is about a factor of 100 lower ($T \simeq 1\,\mathrm{eV}$) in order to produce the spectral break from the neutron decays at the right place, since the neutrinos from neutron decays have about a factor of 100 lower energy than those from the pion decay chain. This scenario is quite exotic, and presumably would require implausible proton luminosities and magnetic fields; we consider these points in detail in Sec.~\ref{sec:discussion}.

The overall situation with respect to both models, with and without magnetic field, is comprehensively contained in the corner plots collected in Figs.~\ref{fig:corner_noB} and~\ref{fig:corner_B}. For the case of small magnetic fields (Fig.~\ref{fig:corner_noB}), the temperature $T$ is constrained from below at $2\sigma$ --- too low temperatures would cause the spectral break move to energies higher than the observed peak of the neutrino spectrum --- and at $1\sigma$ there is a clearly favored range between $80\,\mathrm{eV}$ and $300\,\mathrm{eV}$, coming entirely from the observed spectral break. The other well-constrained parameter is of course the normalization of the diffuse neutrino flux, which is well-measured around the break. On the other hand, for the maximum proton energy $E_{p,\rm max}$ there is only a lower bound --- as we have stressed, the best-fit explanation has basically no upper cutoff and proceeds as a soft power law above the break.

The correlation structure among different parameters also provides significant information. There is clearly a correlation between $E_{p,\rm max}$ and the spectral index $\alpha_p$, which is driven by the aforementioned possibility of explaining the above-break behavior either as a power law or an exponential cutoff. Therefore, harder spectral indices still provide a reasonable fit together with a low $E_{p,\rm max}$, explaining the direct correlation. Notice also that for hard spectral indices, with $\alpha_p<2$, the break may be caused entirely by the low $E_{p,\rm max}$, so the need for a temperature around hundreds of eV disappears. For this reason, at small $\alpha_p$ the range of allowed temperatures widens up considerably, and it can take any value above $T\gtrsim 10\,\mathrm{eV}$, since it is not needed anymore to cause the break. 

For the magnetic-field model (Fig.~\ref{fig:corner_B}), there is a well-defined best-fit solution, whose properties we have already described, but the likelihood shows a clear bimodal structure, with a second minimum corresponding to the scenario D we have highlighted. Overall, there is a clearer preference for a softer proton spectrum, since  a harder proton spectrum could be accommodated  by a low $E_{p,\rm max}$ in the small magnetic field scenario, as we discussed (and $E_{p,\rm max}$ is fixed to a high value here).

As far as  the goodness-of-fit for both models is concerned,  we cannot directly compare the quality from their chi-squared since they are not nested models. 
Nevertheless, the chi-squared still contains qualitative information as to how well the model fits the observations. Defining the chi-squared variable as $\chi^2_\alpha=-2\max{\btheta_\alpha}\log\mathcal{L}(\btheta_\alpha)$, we find for the no-magnetic-field model $\chi^2_{\rm no\,B}=15.5$ and for the magnetic-field model $\chi^2_{\rm B}=13.3$. Considering a number of degrees of freedom equal to 9 (13 flux data points minus 4 parameters) the reduced chi-squared are therefore $1.73$ and $1.48$ respectively. The corresponding p-values, assuming a standard chi-squared distribution with 9 degrees of freedom (this is mostly for qualitative guidance, due to the limited number of degrees of freedom), are $P_{\rm no\, B}\simeq 0.08$ and $P_{\rm B}\simeq 0.15$, so that a statistically significant preference for one or the other cannot be obtained at $2\sigma$ confidence level.

We have also considered how the fit itself would change by including a high-energy point corresponding to the combined KM3NeT-IceCube description of KM3-230213A at the highest energies, above $E_\nu=10^8\,\mathrm{GeV}$~\citep{KM3NeT:2025ccp}. We find that no significant change in the fitted models, and, in particular, Fig.~\ref{fig:diffuse} remains entirely unchanged. The reason is that the model does not allow for a significant spectral hardening at high energies.  Similarly, when fitting the MESE, rather than the combined-fit, data sample, we find that the main qualitative features remain unchanged, with minor quantitative changes: the results of this analysis are reported in Sec.~\ref{sec:mese}.

\section{Predictions for future neutrino telescopes}
\label{sec:predictions}

\begin{figure*}[t]
    \includegraphics[width=\textwidth]{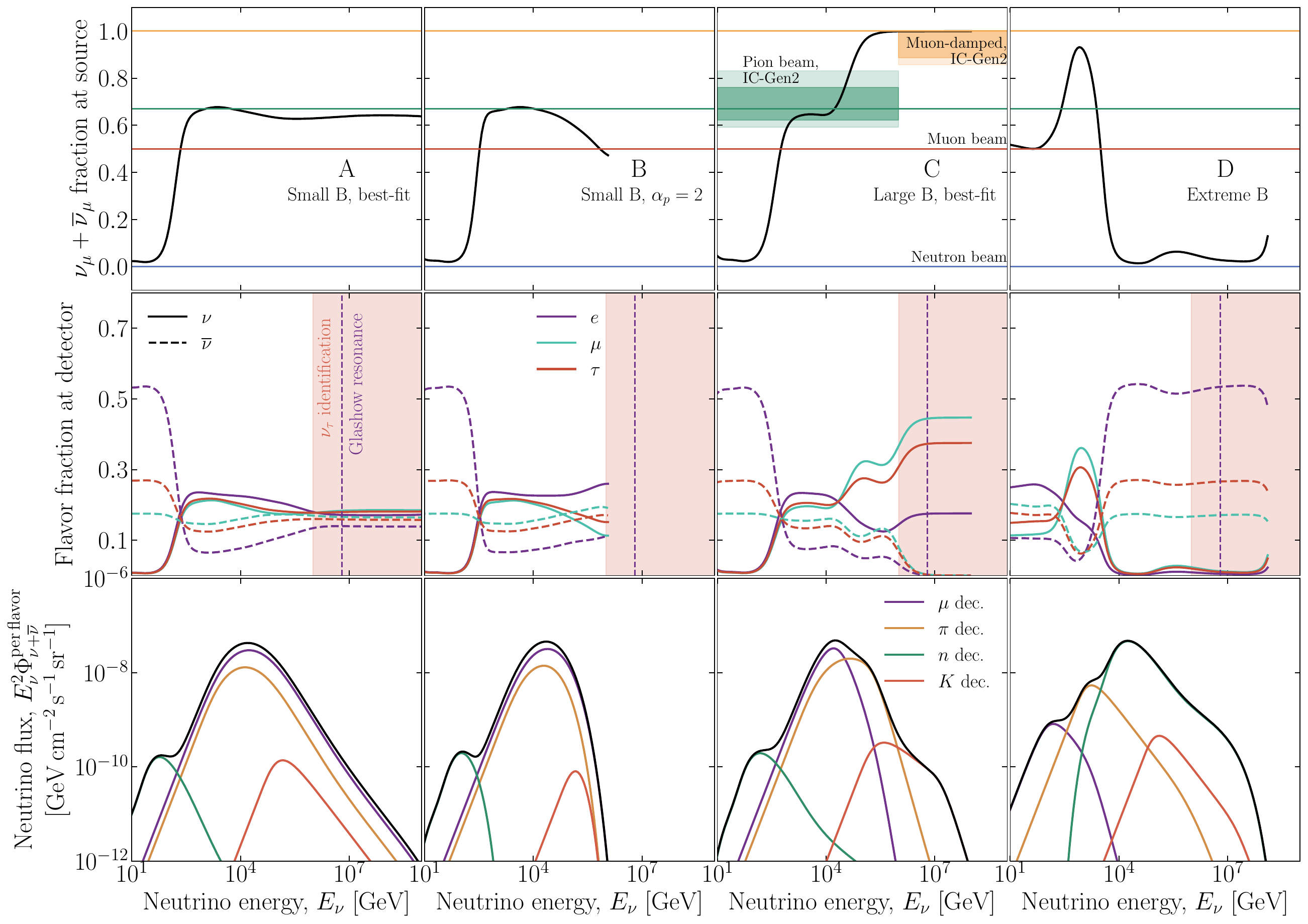}
    \caption{Flavor and neutrino-antineutrino composition at the source (top row) and at detector (middle row), and different parent contributions (bottom row) of the diffuse neutrino flux, for all four benchmark scenarios (in columns). The flavor fraction at the source (top row) is shown in terms of the $\nu_\mu+\overline{\nu}_\mu$ fraction of the whole flux; we also display the reconstruction intervals for pion-beam and muon-damped compositions projected for IceCube-Gen2~\citep{IceCube-Gen2:2020qha}, and the reference values for pion beam, muon-damped, muon beam, and neutron beam compositions. For the flavor fractions at the detector (middle row), we separate the fractions of $\nu_e$, $\nu_\mu$, $\nu_\tau$, $\overline{\nu}_e$, $\overline{\nu}_\mu$, and $\overline{\nu}_\tau$ at Earth (of the total flux), using the flavor mixing parameters $\theta_{12}=33.7^\circ$, $\theta_{13}=8.5^\circ$, $\theta_{23}=48.5^\circ$, $\delta_{\mathrm{CP}}=177^\circ$ to compute neutrino mixings (global fit, normal ordering, from \citet{Esteban:2024eli}). Flavor fractions are shown in the energy range where the flux is at most five orders of magnitude below the maximum only. We also highlight in red shadings the energy range where $\nu_\tau$ detection (corresponding to the red curves for the $\nu_\tau$ fractions) from double-bang events is potentially accessible \citep{Learned:1994wg}, and by a dashed purple line the energy of the Glashow resonance at 6.3~PeV (corresponding to the purple dashed curve for the $\bar\nu_e$ fraction), allowing a determination of the $\overline{\nu}_e$ fraction. In the bottom panels, we separate the neutrino fluxes in terms of their parents (muons, pions, kaons, neutrons), which are differently affected by magnetic field effects. }\label{fig:decomposition}
\end{figure*}

The identification of multiple possible descriptions for the spectral break and shape immediately raises the question whether it is possible to disentangle them based on complementary information, especially from the flavor structure and the fraction of neutrinos and antineutrinos in the diffuse flux. In this section, we consider this question, by highlighting the main differences among the four benchmark scenarios in terms of flavor and neutrino-antineutrino structure.

The overall structure of the neutrino flux is encoded in Fig.~\ref{fig:decomposition}, where we decompose the neutrino emission from decay of muons, pions, neutrons, and kaons for each scenario  (lower row).\footnote{Note that the per-flavor neutrino flux here still corresponds to the all-flavor flux divided by three, which means that all neutrino fluxes from the same parent -- regardless of flavor -- are summed over and divided by three.} The relative dominance of the decay of different secondaries to the neutrino flux leads to a characteristic energy dependence of the flavor structure, also shown in Fig.~\ref{fig:decomposition} in terms of the fractions of different species at source (top row) and detector (middle row), where for the latter a certain choice for the neutrino mixings was made (see figure caption). We now explain for each scenario the physical mechanisms at work in greater detail:

For \textbf{scenario A} (small $B$, soft $\alpha_p=3$), the peak of the neutrino flux is dominated by the typical pion-beam structure, dominated by the pion decay $\pi^\pm\to \mu^\pm + \overset{(-)}{\nu}_\mu$ and subsequent muon decay $\mu^\pm\to e^\pm+\overset{(-)}{\nu}_e+\overset{(-)}{\nu}_\mu$. Note that we take the full $p\gamma$ production modes into account, which means that a substantial fraction of $\pi^-$ will be produced~\citep{Hummer:2010vx}. Since muon decay produces two neutrinos, while pion decay only one, we have the $\mu$ decay contribution roughly equal to twice the $\pi$ decay. At lower energies, a second bump comes from the decay of neutrons, which produces neutrinos with a much lower energy fraction, see next section. The neutrinos from kaon decay, on the other hand, never dominate in this case because the production mode is subdominant. The flavor composition at peak (upper row) directly represents a ``pion beam''.

For \textbf{scenario B} (small $B$, harder $\alpha_p =2$ with proton energy cutoff), the situation is quite similar, except that due to the exponential suppression of the proton flux, the non-thermal power-law tail at high energies is suppressed. Since data do not clearly support the power law hypothesis at high energies at present, scenario B may still be a viable alternative, even though it predicts a lower flux at the energies of \uheevent.

On the other hand, the cases with strong magnetic fields exhibit a completely different phenomenology. For \textbf{scenario C} (large magnetic field), the magnetic field is strong enough to cool the muons, so that immediately above the peak there is a dominance of neutrinos from pion decay (3rd panel, lower row). This regime, usually called the muon-damped regime (see e.g.~\citep{Kashti:2005qa}), has a characteristically different flavor structure, with a somewhat larger fraction of $\nu_\mu$ and $\nu_\tau$ at the detector (3rd panel, middle row), coming from the $\nu_\mu$ dominance at the source from pion and kaon decays (3rd panel, upper row). The difference between the pion and kaon regimes can be seen in the middle row: in the pion beam regime, both $\pi^+$ and $\pi^-$ contribute, leading to a substantial anti-neutrino fraction, whereas in the kaon regime, we only take into account the leading $K^+ \rightarrow \mu^+ + \nu_\mu$ decay mode, which leads to a suppression of the antineutrino fluxes. For a more precise interpretation in that energy range, sub-leading kaon and (charmed) meson channels need to be included, see discussion in \cite{Kachelriess:2006ksy,Kachelriess:2007tr,Lipari:2007su}. 

Finally, \textbf{scenario D} (extreme magnetic field) differs from the previous ones across the entire energy range, due to the extreme magnetic field assumed. In this case, all charged secondaries are synchrotron-cooled to very low energies, so that the dominant peak comes indeed from neutron decay~\citep{Hummer:2010ai}. This means that the peak of the diffuse neutrino flux has a flavor composition completely different from the pion beam one, with a strong dominance of $\overline{\nu}_e$; this regime is usually known as ``neutron beam''. Note that in this case, the flavor composition at peak is dominated by neutron decays, whereas in all other cases, the flavor composition changes into a neutron beam at low energies (see upper row).

The different flavor structures of the various scenarios explaining the spectral break can be exploited to discriminate them. From the observational perspective, the current capability of IceCube in discriminating different flavor compositions is still limited. The main observable that is adopted in this respect is the relative contribution of track and cascade events; track events are primarily driven by charged-current interactions of muon neutrinos, producing a muon which leaves a detectable track of Cherenkov radiation. Neutral-current interactions, as well as charged-current interactions of electron and tau neutrinos, typically produces a cascade-like deposition of energy. At very high energies, the charged-current interaction of tau neutrinos may produce a different topology, with a first cascade at the neutrino interaction vertex spatially and/or temporally separated from a second vertex where the produced tau decays~\citep{IceCube:2021rpz}. This double cascade may provide additional information on the fraction of tau neutrinos in the original neutrino flux; we roughly identify the energy range where this identification would be possible by red regions in Fig.~\ref{fig:decomposition}.

The most recent IceCube reconstruction analysis of the diffuse neutrino flux flavor structure, based on these observables, is~\citet{Abbasi:2025fjc}. Assuming an energy-independent flavor composition, and an equal fraction of neutrinos and antineutrinos, the MESE data sample favors a pion-beam flavor composition, excluding the neutron-beam flavor composition at the 95\% confidence level. The muon-damped flavor composition is only in tension with the measurements at the 68\% confidence level, but not yet excluded. One caveat to this analysis is the assumption of an energy-independent flavor composition; if this assumption is relieved, it is much harder to draw clear conclusions on the flavor composition of the neutrino flux, and at present it is not yet possible to identify clear evidence for a transition between different flavor regimes~\citep{Liu:2023flr}.

From the current flavor measurements, one may conclude that scenario D is already currently excluded, due to its vast predominance of neutron-beam neutrinos across the entire energy range. On the other hand, scenario C, predicting a transition close to the diffuse neutrino flux peak from pion-beam to muon-damped regime, cannot yet be excluded from its flavor structure. Scenarios A and B, with no impact from magnetic field cooling on secondaries, predict a pion-beam composition in quite good agreement with the current IceCube data.

A complementary property, much harder to reconstruct, is the relative fraction of neutrinos and antineutrinos. The main observable in this case comes from the Glashow resonance reaction $\overline{\nu}_e+e^-\to W^-$ at 6.3~PeV (marked by purple dashed lines in Fig.~\ref{fig:decomposition}), with the $W^-$ boson subsequently decaying into hadronic or leptonic channels. IceCube has reported on the detection of one such event~\citep{IceCube:2021rpz} with an estimated cascade energy around $6\,\mathrm{PeV}$. Hence, scenario C, which, due to its strong muon-damped regime, exhibits a near-vanishing fraction of $\overline{\nu}_e$ above 1~PeV in the regime of kaon dominance, might already be in tension with this measurement.\footnote{The null hypothesis that this event comes from (non-resonant) CC or NC interactions of astrophysical neutrinos was calculated (depending on the spectral index) to be between 1.6$\sigma$ and 2.7$\sigma$ in \citet{IceCube:2021rpz}, with 2.3$\sigma$ for a (neutrino) spectral index of 2.5. This tension can be directly interpreted as tension with scenario~C because that scenario predicts zero Glashow events.} On the other hand, a lower bound on the fraction of $\overline{\nu}_e$ in the flux has not been reported by the IceCube collaboration, due to the low statistics in the neutrino flux at the highest energies. 

While the current observational situation has several corners which have not been fully clarified, the combination of future neutrino telescopes with incremented exposure and complementary capabilities may strongly improve it. The uncertainty region on the energy-averaged flavor composition of the neutrino flux should shrink considerably~\citep{Song:2020nfh} with the advent of several upcoming or under-development telescopes, including Baikal-GVD~\citep{Baikal-GVD:2018isr}, IceCube-Gen2~\citep{IceCube-Gen2:2020qha}, KM3NeT~\citep{KM3Net:2016zxf}, P-ONE~\citep{P-ONE:2020ljt}, TAMBO~\citep{Thompson:2023pnl}, and TRIDENT~\citep{Ye:2022vbk}. The IceCube-Gen2 experiment alone would have the capability to clearly differentiate a transition to a muon-damped regime, as we also highlight in Fig.~\ref{fig:decomposition} using the projected reconstruction uncertainties from~\cite{IceCube-Gen2:2020qha}. Even energy-dependent flavor transitions may become in principle distinguishable~\citep{Liu:2023flr}, if they happen in an energy region where the flux has not yet decreased considerably --- this would be the case, for example, for our scenario C. Increased statistics from Glashow resonance events would contribute to constraining these scenarios~\citep{Biehl:2016psj,Liu:2023lxz}, by providing robust lower bounds on the fraction of $\overline{\nu}_e$ in the diffuse neutrino flux. Our obtained fluxes and flavor compositions could serve as benchmarks to test different scenarios in the future.

\begin{figure}[t]
\centering
    \includegraphics[width=\columnwidth]{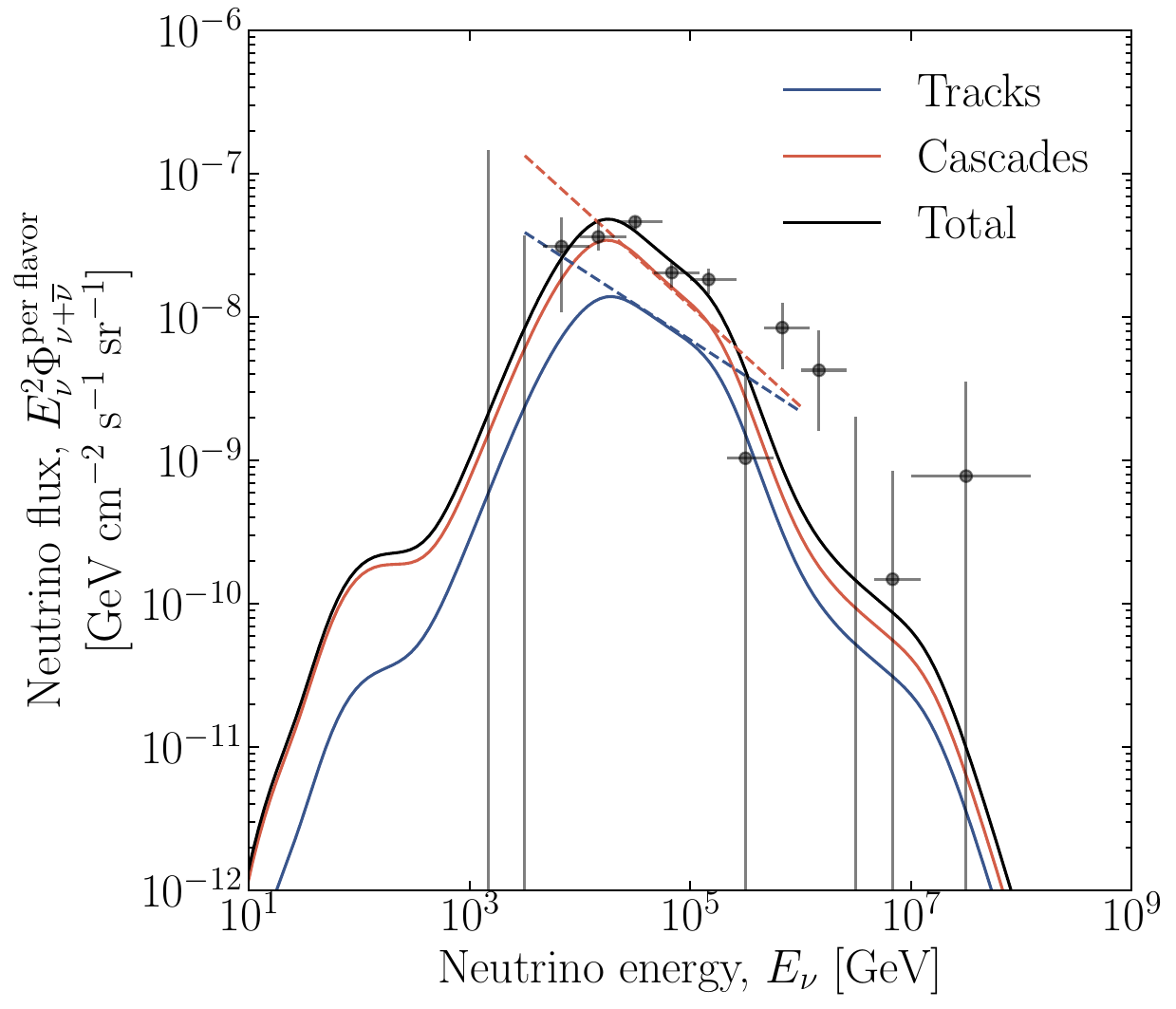}
    \caption{Predicted flux for tracks and cascades separately, for our scenario C. Track and cascade fluxes are obtained assuming that electron and tau neutrinos and antineutrinos always produce cascades, while muon neutrinos and antineutrinos produce tracks with a probability $p_T=0.8$~\citep{Palladino:2015zua} and cascades with a probability $1-p_T=0.2$. The flux is shown per flavor, i.e., divided by a factor of three in consistency with previous figures. We also show as dashed lines a power-law fit for the cascades ($\Phi_{\nu+\overline{\nu}}^{\rm per\;  flavor}\propto E_\nu^{-2.7}$) and the tracks ($\Phi_{\nu+\overline{\nu}}^{\rm per\;  flavor}\propto E_\nu^{-2.5}$).}\label{fig:trackscascades}
\end{figure}

Finally, an interesting feature of scenario C is that it predicts a different spectral shapes for tracks, which are primarily driven by muon neutrinos and antineutrinos, and cascades, mostly driven by the electron and tau flavors. In Fig.~\ref{fig:trackscascades}, we show the predicted spectral shape for the neutrinos contributing to the two topologies, together with a power-law tangent to the relevant energy region between about 10 and 100~TeV: $E_\nu^{-2.5}$ for muon tracks, and $E_\nu^{-2.7}$ for cascades. Intriguingly, cascades exhibit a significantly softer neutrino spectrum compared to tracks. This is especially interesting since IceCube has long time reported a slight tension between the spectral index inferred from tracks and cascades, with the latter pointing to a softer spectrum.  On the other hand, in the comparison of measurements from the IceCube collaboration~\citep{IceCube:2024fxo} (e.g. Fig. 19), the tension seems to be significantly relieved. Nevertheless, we show here that a difference in the inferred spectrum among different neutrino topologies is a potential observable that might in the future help discriminating among different scenarios of neutrino production.

\section{Discussion and limitations}
\label{sec:discussion}

\begin{figure*}[t]
    \includegraphics[width=\textwidth]{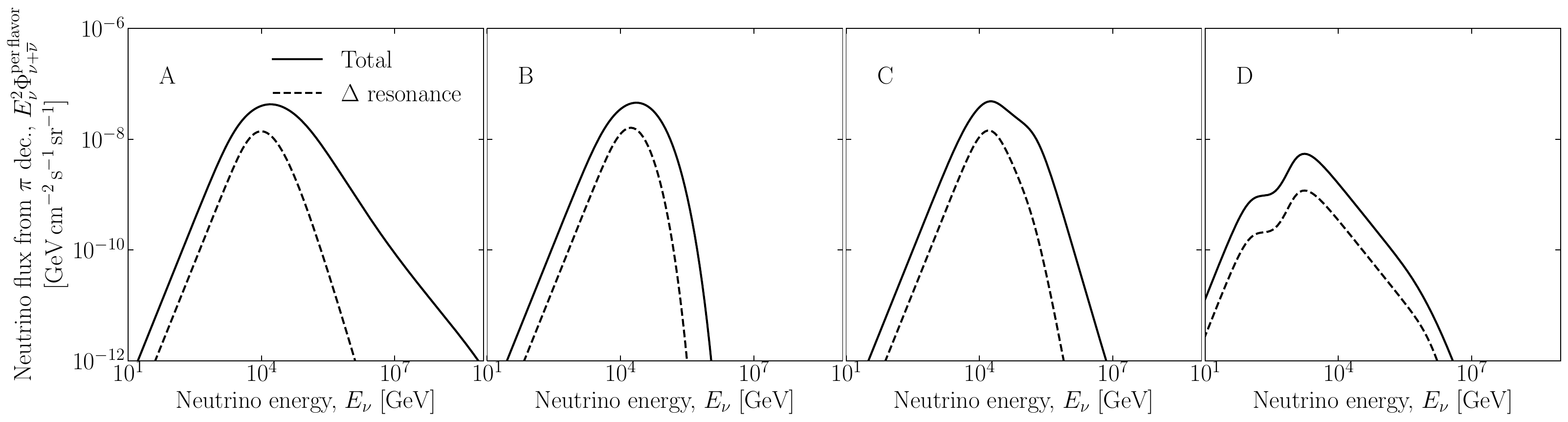}
    \caption{Diffuse neutrino flux and its contribution from the $\Delta$-resonance alone (dashed curve, as defined in model ``Sim-B'' from \citet{Hummer:2010vx}) in comparison to the full photohadronic contribution (solid curves). We only show neutrinos from pion and muon decays.}\label{fig:Delta}
\end{figure*}

{\bf Relationship to  non-thermal target photon models}.
A first important ingredient is the importance of the $\Delta$-resonance contribution $p+\gamma\to \Delta^+\to n+\pi^+$ in the thermal model, especially if power-law spectra are to be described. The $\Delta$-resonance  is often adopted as an approximation for the overall $p\gamma$ emission since it dominates close to the peak of the neutrino flux, which is in many cases a good approximation. In Fig.~\ref{fig:Delta}, we show the separate contribution from the $\Delta$-resonance (model Sim-B in \cite{Hummer:2010vx}) in comparison with the total $p\gamma$ computation. For most scenarios, the $\Delta$-contribution follows closely the total flux (apart from the lower normalization because only a part of the total cross section is taken into account). However, for case A, it fails to reproduce the power-law tail, which in this case is crucial to fit the high-energy neutrino events above hundreds of TeV. This failure comes precisely from the resonant character of the $\Delta$-contribution, which makes the neutrino flux decrease exponentially above the resonance, where there are no target photons exactly satisfying the resonance condition. Instead, the total diffuse neutrino flux continues as a power law $\propto E_\nu^{-\alpha_p}$ (plus small corrections from the energy-dependence of the cross section) driven by the multi-pion contribution, a feature emphasized in~\citet{Fiorillo:2021hty}, which similar to the expectation from $pp$ interactions. Thus, the interpretation of power law neutrino spectra crucially relies on the multi-pion production in the chosen model with thermal target photons. For non-thermal target photon spectra, the neutrino power law index scales as $E_\nu^{-\alpha_p+\beta-1}$, where the target photon differential (in energy) spectrum $\propto \varepsilon^{-\beta}$, see e.g. \citet{Winter:2012xq}. One can easily see that the mapping $\alpha_p \rightarrow \alpha_p - \beta + 1$ between our analysis and the non-thermal target model applies in the high-energy regime, and that the result is degenerate between $\alpha_p$ and $\beta$ then. This especially holds for $\beta < 1$, where the target photon (number) density $\propto \varepsilon^{-\beta+1}$ peaks at high energies. For $\beta>1$, the cutoff shape below the peak around 30~TeV can be slightly affected by $\beta$ (see examples in \citet{Fiorillo:2021hty}), which is at the moment hardly testable due to limited statistics there, but may be interesting the future.

{\bf Interpretation of the normalization.} One limitation of our approach is that it focuses on shape and (flavor and neutrino-antineutrino) composition of the astrophysical flux, whereas the interpretation in terms of a diffuse flux from a population of sources requires additional ingredients, see e. g. \citet{Murase:2013rfa,Murase:2016gly}. In general, the neutrino diffuse flux (for steady sources) is proportional to neutrino luminosity (per source) and local source density $\rho$ as $E_\nu^2 \Phi_\nu \propto L_\nu \,  \rho$, where we omit redshift distribution and bolometric correction factors. The all-flavor neutrino luminosity $L_\nu \simeq 3/8 \,  \tau_{p \gamma} \, \chi_{p \rightarrow \pi} \, L_p$ is proportional to proton injection luminosity $L_p$ and optical thickness $\tau_{p \gamma}$ in the optically thin case (here $\chi_{p \rightarrow \pi} \simeq 0.2$ is the fraction of proton energy going into the pion). Hence, $E_\nu^2 \Phi_\nu \propto \tau_{p \gamma} \times \rho \times L_p$, and there is a degeneracy among these quantities. For example, an abundant source class which is optically thin to $p\gamma$ interactions and has sufficiently high $L_p$ (such as inferred for AGN blazars or low-luminosity GRBs) cannot be distinguished from a rarer source class which is an efficient neutrino producer (such as high-luminosity GRBs) in our approach, and other constraints such as from neutrino multiplets or stacking searches are needed. We can however easily see that in the strong cooling case (scenario D) $L_\nu \simeq 3/2 \,  \tau_{p \gamma} \, \chi_{p \rightarrow n \rightarrow \nu} \, L_p$, where the fraction of energy going into the neutrino from neutron decays $\chi_{p \rightarrow n \rightarrow \nu} \simeq 0.8 \times \, 5 \times 10^{-4}$ \citep{Lipari:2007su} (the neutron takes about 80\% of the proton energy). Thus, since $\chi_{p \rightarrow \pi \rightarrow \nu} \simeq 0.2 \times \, 0.25 \simeq 0.05$, the expected neutrino flux is about a factor of 100 lower (and peaks at a factor of 100 lower energy) than for the pion production chain, and, consequently, a factor of 100 higher proton injection luminosity (in addition to the extreme values of $B$) per source is needed to power the diffuse flux. Thus, apart from a pure neutron beam composition being challenged by flavor data, both energetics and magnetic field for this scenario are challenging. Another limitation of our model is that the proton properties are described by two (effective) parameters -- spectral index and maximal energy -- without addressing the consistency with the environmental properties explicitly: for example, a large magnetic field in the acceleration zone may limit the maximum proton energy via synchrotron losses, 
depending on the balance between cooling and acceleration rate -- which we have not specified. Instead, we assume that the accelerator produces a spectrum with a cutoff energy $E_{p,\mathrm{max}}$ in order to describe the phenomenological properties of the spectral break in the diffuse neutrino flux.

{\bf Optically thick sources.}
Note that we do not explicitly address optically thick sources, but we do not expect large quantitative differences in the spectral shape and flavor composition driven by $\alpha_p$, $E_{p,\mathrm{max}}$, $T$ and the secondary cooling, see e. g. 
App.~C in \citet{Biehl:2017zlw}, since the interaction properties of protons and neutrons are similar. However, we expect that a) there will be suppressed neutrino production from neutron decays, since neutrons will interact before they can decay, b) there will be changes in the neutrino-antineutrino composition where the pion production chain dominates~\citep{Biehl:2016psj} since $\pi^+$ and $\pi^-$ are produced at comparable rates. In addition, the linear dependence on $\tau_{p \gamma}$ is lost, which primarily affects the normalization, but can also introduce a spectral break in the proton spectrum, see Sec.~\ref{sec:methods}. The target photon spectrum can be also modified in sources if the secondary electron/positron or photon injection luminosities from the pion decay chain (which are comparable to the neutrino luminosity) are comparable to the thermal target luminosity -- in which case, the system becomes non-linear. Note that this effect may also occur in optically thin sources if the proton luminosity times optical thickness is high enough, see previous paragraph, but may be more likely in optically thick sources.

{\bf Origin of the spectral break.}
We have confirmed the statistical significance of the spectral break around $20-30\,\mathrm{TeV}$ with a completely independent, self-consistent interaction model for neutrino production. Within this thermal model, the break may come A) from a break in the photohadronic efficiency due to a characteristic photon temperature around $200\,\mathrm{eV}$, or B) from a maximal proton energy around hundreds of TeV, leading to an exponential suppression of the neutrino flux around tens of TeV, or C) from strong magnetic cooling of secondary particles, requiring magnetic fields around $B\sim 60\,\mathrm{kG}$. The fourth scenario that we consider is based on D) dominant neutrinos from neutron decay, with all other components suppressed by an extreme magnetic field. Such a scenario is already in tension with the current flavor composition measured by IceCube and the increases proton luminosity requirement, as we have discussed earlier. There is a fifth possibility, which our model, with a single power-law proton spectrum, by construction does not accommodate: the photohadronic efficiency has no break, but E) the proton spectrum itself may be a broken power law, with a break precisely at hundreds of TeV. In this case, the spectral break would imprint itself onto the neutrino spectrum.
 Note, however, a neutrino spectrum from a broken power law proton spectrum can be effectively described by our model in scenario A) if the neutrino spectral break originates from the proton break and the target photon spectrum is a simple, featureless (non-thermal) power law: one would choose an effective $T_{\mathrm{eff}} \simeq 67 \, \mathrm{eV} \times  \mathrm{PeV}/E_{p,\mathrm{break}}$ from \equ{ET} (where $E_{\nu,\mathrm{thr}} \rightarrow 0.05 \, E_{p,\mathrm{break}}$) to simulate the proton spectral break. Using \citet{Fiorillo:2023dts} as an example, one obtains $T_{\mathrm{eff}} \simeq 1 \, \mathrm{keV}$ for $E_{p,\mathrm{break}} \simeq 60 \, \mathrm{TeV}$ ($E_{p,\mathrm{cool}}$ therein), together with the same post-break spectral index $\alpha_p \simeq 3$ obtained there (which corresponds to the neutrino spectral index in that model beyond the cooling break). These inferred parameters are roughly consistent with our fit, but the spectral break is at very low energies in comparison to diffuse flux data. This is not surprising, since neutrinos from NGC~1068, to which the coronal model was tuned, peak significantly below the spectral break of the diffuse neutrino flux. Conversely, the best-fit in Tab.~\ref{tab:models} can be translated into $E_{p,\mathrm{break}} \simeq 300 \, \mathrm{TeV}$ by this formula, which corresponds to the ``typical break energy'' expected for the population.  The spectral index below the break cannot be directly controlled in the thermal model similar to the case of non-thermal target photons discussed earlier. Note that $T_{\mathrm{eff}}$ is an effective temperature here, which is not related to the target photons: The photohadronic interactions are dominated by the low target energies here ($\beta>1$ for the coronal X-rays, see earlier discussion for non-thermal targets) with a sufficiently low cutoff energy $E_{X,\mathrm{min}} \ll 3 \, T_{\mathrm{eff}}$. Thus, the neutrino spectrum from scenario E) can be effectively reproduced within the framework of scenario A); the reason is that in both cases the neutrino spectrum around the peak is a broken power law, with the break imprinted from the photohadronic efficiency in scenario A), while inherited from the proton spectral break in scenario E).  

{\bf Relationship to astrophysical source models.} It is interesting that one or several of the scenarios above appear in plausible astrophysical source models. A strong candidate source for neutrino emission around the spectral break is the class of Seyfert galaxies, emerging as increasingly correlated with significant excesses of events in the neutrino sky~\citep{IceCube:2024dou,Abbasi:2025tas,IceCube:2026hzq}. The brightest neutrino hotspot in the sky coincides with the Seyfert galaxy NGC~1068, with an excess of events around a few TeV~\citep{IceCube:2022der}, further corroborating a scenario where these sources contribute to the diffuse emission. The most plausible emission site of neutrinos from Seyfert galaxy is the corona, a compact region close to the central supermassive black hole of the galaxy which naturally agrees with the lack of associated high-energy gamma-ray emission: we refer to~\cite{Padovani:2024ibi} for a recent review. The current models for coronal neutrino emission, intriguingly either assume scenario B), with a maximal proton energy from a balance of cooling and acceleration roughly around hundreds of TeV~\citep{Murase:2019vdl,Inoue:2019yfs,Fiorillo:2024akm}, which is naturally met in scenarios with a magnetized turbulent corona, or scenario E), with a proton spectral break around $100$~TeV caused by acceleration in a magnetic reconnection layer within the black hole magnetosphere~\citep{Fiorillo:2023dts}. Intriguingly, \cite{Karavola:2024uui,Karavola:2026rpg} have recently pointed out that in the latter scenario the strong magnetic fields around tens of kG also naturally realize scenario C), with a strong muon damping around the break of the neutrino spectrum -- while the maximal proton energy is self-consistently determined within the acceleration model, including proton synchrotron losses. Therefore, the mechanisms we have identified here seem to directly arise in the coronal models of Seyfert galaxies, and indeed both in the turbulent scenario~\citep{Murase:2019vdl,Padovani:2024tgx,Ambrosone:2024zrf,Fiorillo:2025ehn,Murase:2026hrz,Yang:2026syv} and in the magnetized reconnection scenario~\citep{Karavola:2026rpg} they have been proposed as dominant contributors to the diffuse neutrino flux. Other possible source classes exhibiting substantial secondary cooling (scenario C) may be choked or low-luminosity Gamma-Ray Bursts \citep{Senno:2015tsn,Boncioli:2018lrv,Carpio:2020app}, where the neutrino spectral break energies driven by the break of the non-thermal target photons, however, often occur at higher energies.  Note again that scenario E) and be identified with scenario A) using an effective parameter mapping, as discussed earlier.

{\bf Additional population at highest energies?} Describing the spectral break and the high-energy neutrino flux within a single-population framework does not, of course, exclude the existence of an additional population at high energies. Incorporating two populations within the thermal model would substantially enlarge the parameter space, yielding limited statistical benefit at this stage of phenomenological analysis. For example, one may duplicate the small $B$ model and add a self-similar contribution with hard $\alpha_p \simeq 1$ and $E_{p,\mathrm{max}} \simeq 10^{11} \, \mathrm{GeV}$ which leads to a spectral hardening at a few PeV and describes the KM3NeT data point. While such a model slightly improves the fit, the goodness of fit decreases substantially because of the increased number of parameters. Therefore, such a contribution is at present not statistically significant (nor can it be excluded) even if the KM3NeT event is taken into account.
In astrophysics-informed scenarios tied to specific source classes, however, the introduction of a second population may become necessary due to source-specific constraints. For instance, most coronal models struggle to accommodate neutrino production up to PeV energies (see however~\cite{Yang:2026syv}), motivating two-population interpretations that include a high-energy blazar contribution~\citep{Padovani:2024tgx,Karavola:2026rpg,Murase:2026hrz}. Scenarios in which blazars account primarily for the PeV neutrino flux, with other sources dominating the lower-energy regime, have indeed been explored extensively~\citep{Palladino:2018evm,Palladino:2018lov,Ambrosone:2020evo,Padovani:2024tgx,Karavola:2026rpg,Murase:2026hrz}. The transition to such a high-energy bump component could be tested with future observations combining the exposure of multiple neutrino telescopes~\citep{Fiorillo:2022rft}.

{\bf Multi-messenger constraints.} Finally, we briefly comment on multimessenger constraints on this scenario, primarily coming from the diffuse gamma-ray background. Gamma-ray transparent sources are generally excluded from dominating the neutrino flux at tens of TeV, since the associated gamma-ray emission would overshoot the Fermi-LAT observations~\citep{Murase:2015xka}. This already excludes otherwise promising candidates, such as starburst galaxies~\citep{Tamborra:2014xia,Bechtol:2015uqb,Palladino:2018evm,Peretti:2019vsj,Ambrosone:2020evo}, and generally hints at photohadronic sources in which the photon target for $p\gamma$ interactions naturally acts as an absorption target for high-energy gamma rays~\citep{Murase:2015xka}. Our phenomenological thermal model falls indeed within this qualitative class, since it is based on $p\gamma$ neutrino production, although we do not make any specific assumption on the radiative compactness of the source. On the other hand, the flux above 100~TeV can safely be explained by $pp$ sources~\citep{Murase:2013rfa}, and more generally by gamma-ray transparent sources, without tension with the Fermi-LAT measurements.

\section{Summary and conclusions}
\label{sec:summary}

We have assumed that a single source class dominates the diffuse astrophysical TeV--PeV neutrino flux observed by the IceCube experiment. Since the latest data indicate a spectral break at about 30 TeV, we have assumed that this may originate from photohadronic interactions. Based on a recent work which demonstrated that even non-thermal target photons can be approximated by a thermal spectrum from the neutrino spectrum perspective, we have proposed a generic neutrino production model based on four parameters: $\alpha_p$ (spectral index protons), $E_{p,\mathrm{max}}$ (maximal proton energy), $T$ (temperature of a thermal target photon spectrum), and $B$ (magnetic field), where the main effect of $B$ in our model are magnetic field effects on the secondary muons, pions, and kaons. Note that in comparison to frequently used fit models for the spectral shape (such as broken power law or log parabola), our neutrino spectral shape and flavor composition self-consistently arise from photohadronic interactions and secondary cooling. Our approach therefore fills the gap between purely empirical fit models and astrophysical source models -- which typically have a broader scope in terms of multi-messenger predictions, but typically require more ingredients and assumptions.

From combined multi-parameter fit, we have found four scenarios which can well describe the diffuse flux:

{\bf (A) Soft injection spectrum.} We find a soft ($\alpha_p \gtrsim 2.6$) injection spectrum, paired with a high enough $E_{p,\mathrm{max}} \gtrsim 10^7 \, \mathrm{GeV}$ and an effective target temperature $T \sim 80 - 300 \, \mathrm{eV}$ (soft X-rays), which reproduces the newly observed spectral break. Magnetic field effects on the secondaries are negligible if $B \lesssim 10 \, \mathrm{G}$.

{\bf (B) Hard injection spectrum with a proton energy cutoff.} An alternative solution for $\alpha_p \simeq 2$, as expected in shock acceleration mechanisms, requires a similar target photon temperature to produce the spectral break, and a proton energy cutoff $E_{p,\mathrm{max}} \sim 500 \, \mathrm{TeV} -  2 \,\mathrm{PeV}$; here, again, $B \lesssim 10 \, \mathrm{G}$.

{\bf (C) Large $\boldsymbol{B}$ to produce a secondary cooling cutoff.} This solution requires large $E_{p,\mathrm{max}} \gg 10^9 \, \mathrm{GeV}$ and exhibits similar parameter values for $T$ to produce the spectral break. The cutoff is driven by strong magnetic fields $B \simeq 20 - 100 \, \mathrm{kG}$ on the secondary muons, pions and kaons, whereas $\alpha_p \gtrsim 2.2$.

{\bf (D) Extreme values of $\boldsymbol{B}$.} For extreme values of $B \gtrsim 1 \, \mathrm{MG}$, the neutrino flux will be dominated by neutron decays -- whereas the pion and kaon decay chains are completely suppressed. Here a factor of hundred lower $ T \sim 1 \, \mathrm{eV}$ (infrared--optical range) is required to reproduce the spectral break because the neutrons take about a factor of 100 lower fraction of the proton energy. Since this also affects the energy budget required (about a factor of 100 larger proton injection luminosity required) and a pure neutron beam is already challenged by current IceCube flavor composition data, we have considered this possibility exotic -- although such extreme values of $B$ may occur on the vicinity of neutron stars (pulsars).

Comparing our generic results with actual astrophysical models, we have demonstrated that especially scenarios B) and C) share many recently pointed out features for models of Seyfert galaxies. For example, magnetic fields as inferred in our scenario C) are consistent with those found in recent models proposing a magnetic reconnection layer within the black hole magnetosphere. Scenario C) may be also realized in other source classes with strong magnetic fields, such as low-luminosity or choked gamma-ray bursts, whereas scenario B)  applies to sources with a classically proposed $E^{-2}$ acceleration spectrum. Scenario A) -- softer injection spectra following the proton injection spectrum -- has  been traditionally more discussed in the context of starburst galaxies, where $pp$ interactions dominate the neutrino production. Thanks to multi-pion processes, such as scenario can be also found for $p\gamma$ interactions. Scenario A) can be also used to describe a break in the proton spectrum similar to some models for Seyfert galaxies, using an effective parameter mapping.

While the association with a particular astrophysical source class has to rely on different techniques, such as directional correlations or stacking, we expect that the particle physics properties of the neutrino flux can provide further information in the future:
\begin{description}
    \item[Flavor composition] Especially scenario C) (large magnetic field) exhibits a clear flavor transition to a muon-damped source around the spectral break, which should be detectable with increased statistics e.g. of IceCube-Gen2.  Scenario D) exhibits a neutron-beam composition.
    \item[Neutrino-antineutrino composition] The Glashow resonance at 6.3~PeV may allow for a discrimination among scenarios A), B)  versus C) ($\bar\nu_e$ suppressed) or D) ($\bar\nu_e$ enhanced).
\end{description}
As far as the spectral shape is concerned, we cannot uniquely establish a soft power law at the highest energies (scenario A versus B), as it has been frequently suggested by spectral fits of the IceCube collaboration in the past. Similarly, the statistics from the isolated event \uheevent\ is neither strong enough to statistically affect our conclusions, nor to establish an upturn (spectral hardening) of the flux at the highest energies. We however cannot exclude the existence of another flux from a different population of sources at the highest energies.

We conclude that the observation of a spectral break in the diffuse astrophysical neutrino flux is interesting from the perspective of $p\gamma$ neutrino production models, as it can naturally arise from the properties of the target photon spectrum. On the other hand, a $pp$ interaction model requires a low-energy proton cutoff around 30~TeV with an appropriate shape, which are are two properties which are more difficult to motivate physically. Future instruments, such as IceCube-Gen2, will provide not only better statistics on the spectral shape (such as the break and the high-energy cutoff), but also measure the flavor composition as a smoking gun signature of strong magnetic fields in the sources.

\section*{Acknowledgements} 

We would like to thank S. I. Stathopoulos for useful comments on the manuscript.
D.F.G.F. was supported by the Alexander von Humboldt Foundation (Germany), and was hosted by DESY, Zeuthen, in the early stages of this project. This work was supported by the European Research Council, ERC Starting grant \emph{MessMapp}, S.B. Principal Investigator, under contract no. 949555.

\section*{Data availability} 

Neutrino flux predictions (all-flavor fluxes, small or large $B$ models) can be obtained from \citet{winter_2026_19820788}, as well as more detailed flavor- and composition-split fluxes for scenarios A) to D).

\appendix
\section{Results for MESE data sample}
\label{sec:mese}

\begin{figure*}[t]
    \includegraphics[width=\textwidth]{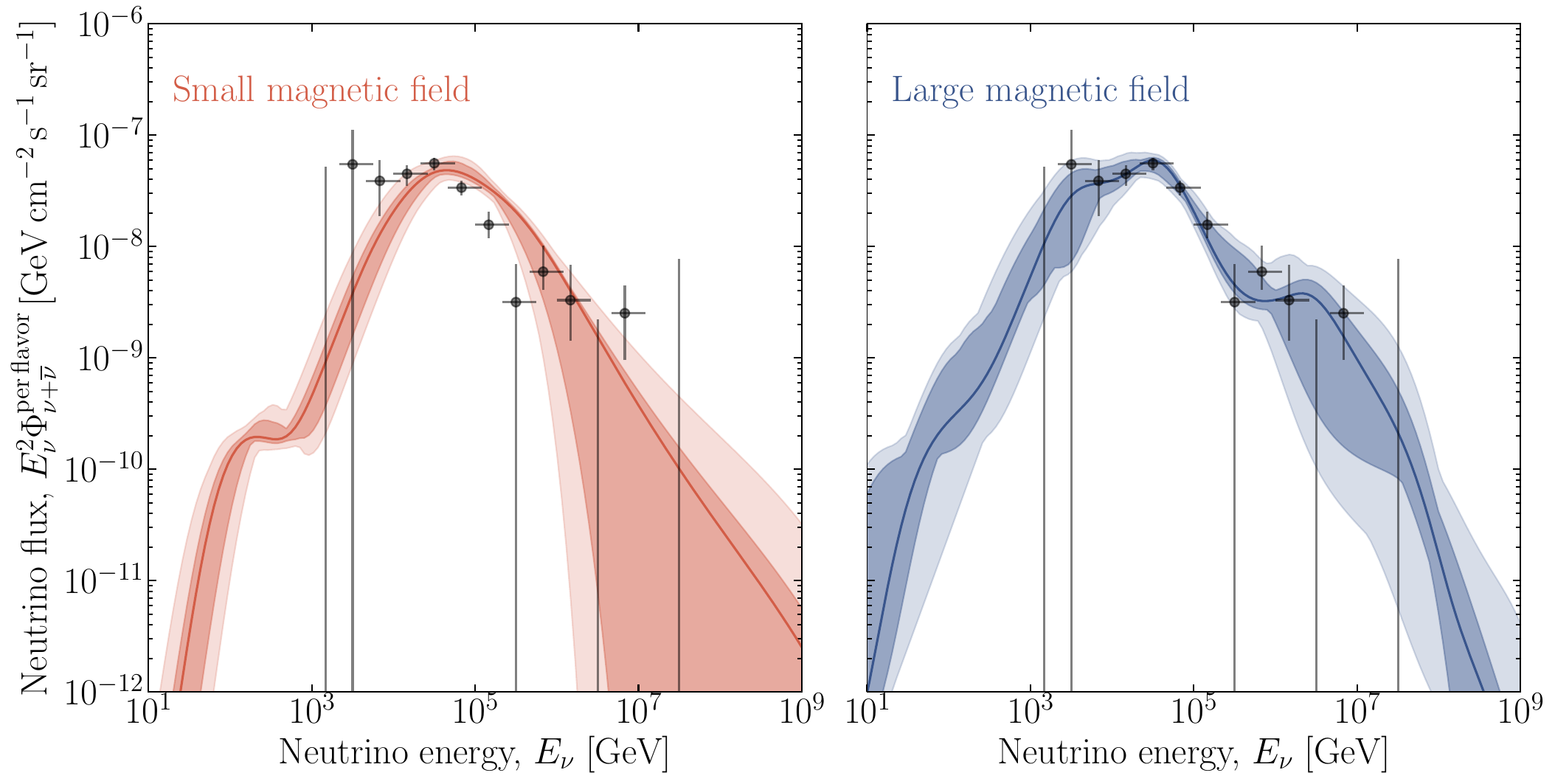}
    \caption{Same as Fig.~\ref{fig:diffuse} in the main text, using the MESE data from Fig.3 of~\cite{Abbasi:2025tas}. We report for this case only the best-fit curves, both for small (left) and for large (right) magnetic fields.}\label{fig:diffuse_mese}
\end{figure*}

In this appendix, we consider how the results of our analysis change by using the segmented flux fit from~\cite{Abbasi:2025tas} rather than the combined fit. Fig.~\ref{fig:diffuse_mese} shows the flux points corresponding to this data set, to be compared with Fig.~\ref{fig:diffuse} in the main text.

The overall tendency of the MESE segmented flux fit is an increased neutrino flux in the low-energy region. In particular, the second energy bin corresponds to a measured flux here, different from the combined fit sample in which there was only an upper bound. This shows up in the comparison with our models, especially in the cases with small magnetic fields. Here, the spectral break induced by the target photon temperature leads to a rather hard spectrum below the break; for a thermal photon spectrum, the photohadronic efficiency changes exponentially at the threshold energy. Therefore, the best-fit spectrum is slightly below the IceCube measurements, although not at a statistically significant level. More importantly, this feature descends entirely from our use of a perfect blackbody spectrum; a power-law tail in the target photon spectrum at high energies would offer enhanced photohadronic efficiency at low proton energies and increase the neutrino flux in this range. 

For the solutions with large magnetic field, on the other hand, the break is caused by the magnetic field itself, so the photohadronic efficiency needs not exhibit this break. This allows for an explanation which captures more precisely the data. Overall, however, the main features of the model comparison with the data are very mildly affected, compared to the combined fit shown in the main text, if we use the MESE sample.

\begin{figure*}[tp]
    \includegraphics[width=\textwidth]{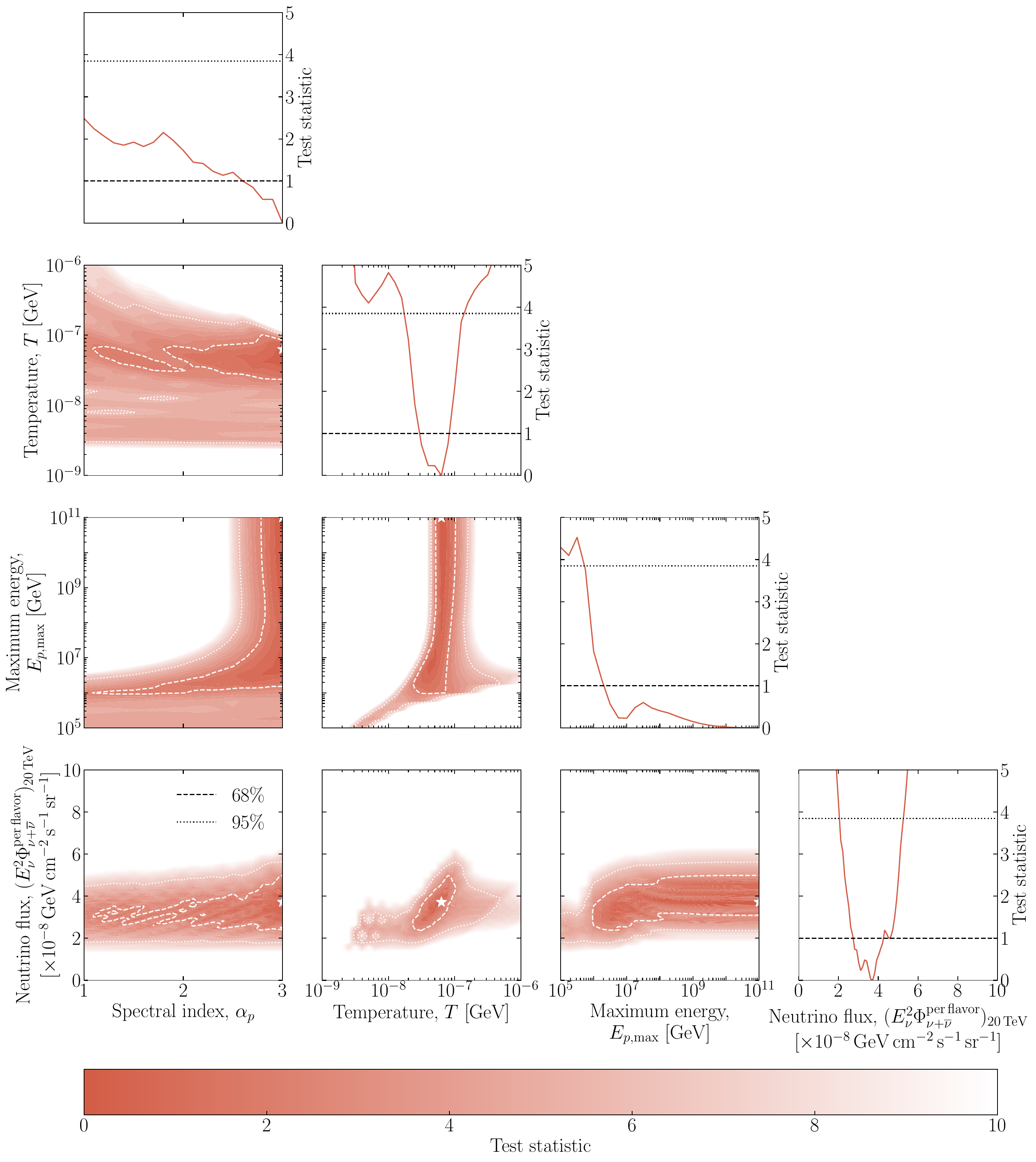}
    \caption{Same as Fig.~\ref{fig:corner_noB}, fitting the MESE data rather than the combined fit data.}
\end{figure*}

\begin{figure*}[tp]
    \includegraphics[width=\textwidth]{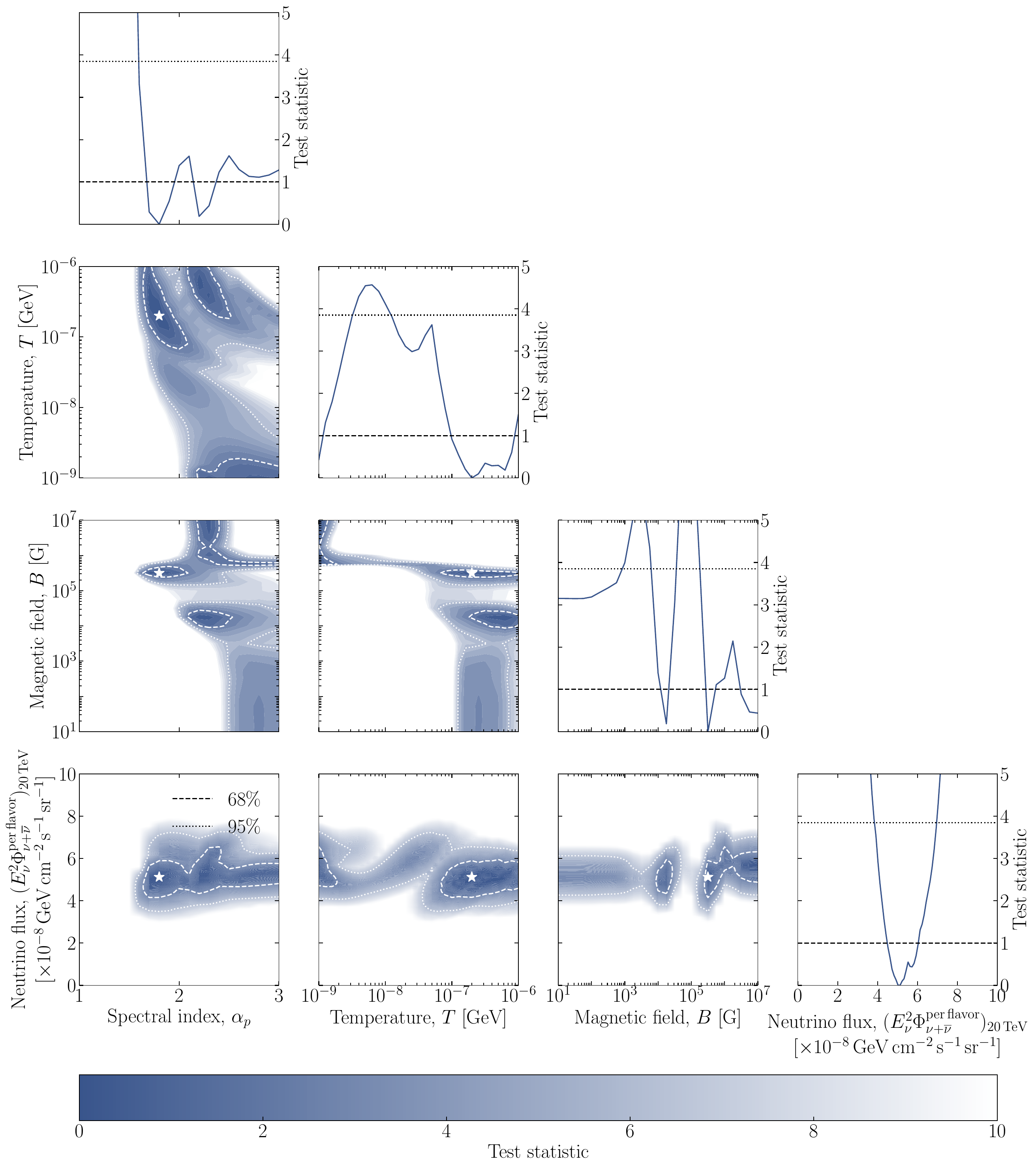}
    \caption{Same as Fig.~\ref{fig:corner_B}, fitting the MESE data rather than the combined fit data.}
\end{figure*}

This shows up also in the corner plots, shown in Fig.~\ref{fig:corner_noB} for the small magnetic field model, and in Fig.~\ref{fig:corner_B} for the large magnetic field model. In the model with small magnetic field, the statistical preference for a temperature around $T\sim 100\,\mathrm{eV}$ is even stronger, due to the prominent spectral break. The favored ranges for all parameters are all compatible with the combined fit in Fig.~\ref{fig:corner_noB}.

The case with large magnetic field is somewhat more interesting, generally favoring harder proton spectra --- the best-fit solution has $\alpha_p<2$ --- with stronger magnetic fields above $B\sim 10^5\,\mathrm{G}$. On the other hand, the bimodal structure of the likelihood also shows up rather clearly, with an additional maximum for the likelihood having smaller magnetic field $B  \sim 10^4\,\mathrm{G}$ and softer proton injection spectrum $\alpha_p\gtrsim 2$. Within the statistics of current IceCube data, there are also no qualitative deviations from the results obtained using the combined fit sample.

\end{document}